# OFDM DEMODULATION USING VIRTUAL TIME REVERSAL PROCESSING IN UNDERWATER ACOUSTIC COMMUNICATION*

YANLING YIN, SONGZUO LIU*, GANG QIAO

*College of Underwater Acoustic Engineering, Harbin Engineering University*
*Harbin, Heilongjiang 150001, China*
*yinyanling@hrbeu.edu.cn*
*liusongzuo@hrbeu.edu.cn*
*qiaogang@hrbeu.edu.cn*
http://www.hrbeu.edu.cn

YUE YANG

Dep. of Electrical Engineering, University of Washington
Seattle, WA, 98105, USA
yueyang@uw.edu

The extremely long underwater channel delay spread causes severe inter-symbol interference (ISI) for underwater acoustic communications. Passive time reversal processing (PTRP) can effectively reduce the channel time dispersion in a simple way via convolving the received packet with a time reversed probe signal. However the probe signal itself may introduce extra noise and interference (self-correlation of the probe signal). In this paper, we propose a virtual time reversal processing (VTRP) for single input single output (SISO) Orthogonal Frequency Division Multiplexing (OFDM) systems. It convolves the received packet with the reversed estimated channel, instead of the probe signal to reduce the interference. Two sparse channel estimation methods, matching pursuit (MP), and basis pursuit de-noising (BPDN), are adopted to estimate the channel impulse response (CIR). We compare the performance of VTRP with the PTRP and without any time reversal processing through MATLAB simulations and the pool experiments. The results reveal that VTRP has outstanding performance over time-invariant channels.

## 1. Introduction

The underwater acoustic channel continues to present significant challenges to robust underwater acoustic communications. Orthogonal Frequency Diversity Multiplexing (OFDM) is an attractive candidate for such communication, due to its high spectrum efficiency, resistance to frequency selective fading, simple channel equalization, and computationally efficient modulation/demodulation [1]-[4]. OFDM eliminates inter-symbol interference (ISI) and inter-carrier interference (ICI) by simply adding a cyclic prefix/postfix. At the same time the underwater channel delay spread usually lasts for tens or even hundreds of milliseconds due to the extremely low speed of sound underwater (1,500m/s). This poses challenges for equalizer design[7][8]. Although the cyclic prefix can eliminate ICI, the selective and deep fading in frequency domain caused by multipath is still a major problem in underwater OFDM systems.

Time reversal or phase conjunction has aroused much interest in underwater communications for its potential in focusing signals at a desired point in a waveguide with high energy and low distortion. Time reversal can be classified into two categories: active time reversal (ATR) and passive time reversal (PTR). ATR uses two-way transmissions to focus the signal which is not suitable for communications[9][10]. PTR is easily embedded in communication systems, in which case, time reversal is done only via one way transmit. The source sends a probe signal (PS) prior to the data packet, and then convolves the packet with the time reversed probe signal, which acts as a backward transmission[11].PTR compresses the distorted signal in time domain and collects all multipath energy to enhance the signal-to-noise ratio (SNR). Time reversal mirror (TRM) was subsequently proposed using an array of transmitters/receivers to focus sound[12]. Taking advantage of the array, the signals were compressed in space and the spatial focus mitigates the effects of channel fading and also provides a high SNR.

PTR/PTRM has been extensively applied in underwater acoustic communications and successfully tested at sea. Authors in [13][14][17] validated the PTRM performance using lake experiments in single carrier communication including BPSK and QPSK modulation. [18] studied a TRM-OFDM system with a simple channel estimation procedure based on the Wiener filter formulation [15][16]. Multiple transmitter and a single receiver were used in the system. [19] used passive time reversal as a multichannel combining preprocessor for impulse response shortening. In the experiment, a 16-hydrophone receiver array was used. The experimental results showed that it was more reliable than maximal ratio combining OFDM (MRC-OFDM).This approach was then improved upon by adopting basis pursuit for identification of sparse responses from PTR channel probes[20].

Prior work shows the remarkable performance of PTRM in OFDM underwater communications. However, dense transmitter/receiver arrays reduce the flexibility and restrict the application. Considering underwater acoustic networks, we adopt a single transmitter single receiver system in this paper to make it more easily connected into the network. For PTR, the performance is limited by self-generated inter-symbol interference caused by the probe signal. Also, the received probe is contaminated by noise which introduces more random interference. In this paper, we propose a method named as virtual time reversal (VTR) which uses the time reversed estimated channel impulse response (CIR) convolved with the received packet to reduce the interference caused by the probe signal. In this work, we adopt two sparse channel estimation methods: matching pursuit (MP)[23][24] and basis pursuit de-noising (BPDN)[25][26][22]to estimate channel response using the probes. We use the frame synchronization - linear frequency modulated (LFM) signal as the probe which can effectively utilize the transmission efficiency and track the channel. We compare the virtual time reversal processing (VTRP) with the passive time reversal processing (PTRP) and without any time reversal processing through MATLAB simulations and pool experiments. The results show that the proposed VTRP could obtain more processing gain benefits from channel estimation.

The rest of this paper is organized as follows. Section 2 introduces the system model including the format of transmitted signal, received signal, channel model and the packet structure. Section 3 presents the three types of receiver processing procedure as well as the sparse channel estimation algorithms. Section 4 and 5 demonstrate the performance of the proposed scheme using numerical and experimental results. Section 6 concludes the paper.

## 2. System Model

### 2.1 Transmitted Signal

Cyclic prefix (CP) OFDM is considered in the system. Let $T$ denotes the OFDM symbol duration. $T_{pre}$ and $T_{post}$ denote the cyclic prefix and cyclic postfix durations. We need to emphasize that it is necessary and important to add a cyclic postfix in the time reversal system. We will discuss this further in Section 3. The total OFDM block duration is $T' = T + T_{pre} + T_{post}$. The frequency of the $k$th subcarrier is

$$f_k = f_c + k/T, \quad k = -K/2,...,K/2-1, \tag{1}$$

where $f_c$ is the carrier frequency and $K$ is the total number of subcarriers. The subcarrier carrier spacing is $1/T$, so the bandwidth is $B=K/T$. Let $d[k]$ and $p[l]$ denote the data information and pilot to be transmitted on the $k$th and $l$th subcarriers, respectively. Pilots are evenly spaced and uniformly distributed over the whole bandwidth. Assume the assemble of the data subcarrier is $S_D$ and the pilot subcarrier is $S_P$, and they satisfy $S_D \cup S_P = \{-K/2,...,K/2-1\}$. The transmitted passband signal is then

$$s(t) = \text{Re}\left\{\sum_{k \subset S_D} s[k]e^{j2\pi f_k t}q[t] + \sum_{l \subset S_P} p[l]e^{j2\pi f_l t}q[t]\right\} \quad t \in [0,T] \tag{2}$$

where $q(t)$ is the pulse shaping filter. In this paper we use a rectangular pulse shaping filter

$$q(t) = \begin{cases} 1, & t \in [0,T] \\ 0, & otherwise \end{cases} \tag{3}$$

### 2.2 Channel Model

The time-varying underwater acoustic multipath channel, consisted of $N_p$ discrete paths, can be expressed as

$$h(\tau,t) = \sum_{p=1}^{N_p} A_p(t)\delta(\tau - \tau_p(t)) \tag{4}$$

where $A_p(t)$ and $\tau_p(t)$ are the amplitude and delay of the $p$th path. In order to simplify the problem, we assume that, with a period (one or several symbol durations), i) the path amplitudes do not change, i.e, $A_p(t) \approx A_p$, and ii) the path delays are linearly changed, i.e.,

$$\tau_p(t) = \tau_p - a_p t \tag{5}$$

where $a_p$ is the Doppler rate corresponding to the $p$th path. So the simplified channel model is

$$h(\tau,t) = \sum_{p=1}^{N_p} A_p \delta\left(\tau - (\tau_p - a_p t)\right) \tag{6}$$

## 2.3 Received signal

According to the transmitted signal and the channel, the received passband signal can be expressed as

$$r(t) = \sum_{p=1}^{N_p} A_p s\left((1+a_p)t - \tau_p\right) + v(t) \tag{7}$$

where $v(t)$ is the additive complex noise. If the receiver can completely remove the Doppler shift, the signal can be treated as passing through an LTI system. The received signal after Doppler compensation can be written as

$$\tilde{r}(t) = s(t) * \tilde{h}(t) + \tilde{v}(t) \tag{8}$$

where * denotes convolution. $\tilde{v}(t)$ is the noise after Doppler compensation and

$$\tilde{h}(\tau) = \sum_{p=1}^{N_p} \tilde{A}_p \delta(\tau - \tilde{\tau}_p) \tag{9}$$

where

$$\tilde{A}_p = \frac{A_p}{1+a_p}, \quad \tilde{\tau}_p = \frac{\tau_p}{1+a_p} \tag{10}$$

## 2.4 Packet structure

The transmitted data packet is composed of several frames. The number of frames depends on the packet size. The frame structure is shown in Fig. 1. The frame consists of three parts: frame synchronization signal (LFM), Doppler estimation signal (CW), and information signal (OFDM). At the beginning of the frame, a liner frequency modulated (LFM) signal is used for frame synchronization. This is then followed by a cosine wave (CW) pulse signal to estimate the Doppler rate. The OFDM blocks are transmitted at the end. A frame may contain several OFDM blocks and the number depends on the channel coherence time. There is a guard interval between every two signals and the guard interval is longer than the maximum multipath channel delay.

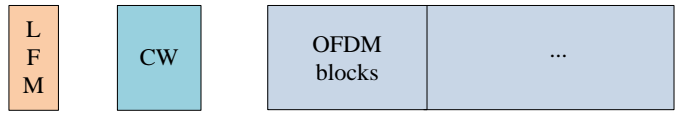

Fig. 1: Frame structure.

## 3 Receiver Processing

In this section, first, we introduce the receiver structure without any time reversal processing. Then, we outline the procedures of passive and virtual time reversal processing. MP and BPDN sparse channel estimation algorithms are introduced in the last part.

### 3.1 Receiver without time reversal processing

The simplified receiver diagram without time reversal processing is shown in Fig. 2. The sampled received signal passes through the following key steps:
1) Synchronization: Synchronize the received signal via correlating the received signal with the transmitted LFM signal. Once the peak of correlation result exceeds the predefined threshold, it indicates a frame arrival. Synchronize the frame with the path which has the strongest power.
2) Extract CW and Doppler rate estimation: After synchronization, extract the CW pulse signal and estimate the Doppler rate. Doppler rate indicates how the received signal has been compressed or dilated by the channel. It can be estimated by comparing the frequency of the received signal $\hat{f}_{rx}$ with the frequency of the transmitted signal $f_{tx}$.

$$\hat{f}_{rx} = (1+\hat{a})f_{tx} \Rightarrow \hat{a} = \hat{f}_{rx}/f_{tx} - 1 \qquad (11)$$

3) Resampling: Resample the frame using the estimated Doppler rate.
4) Channel estimation and equalization: After FFT implementation, estimate the frequency response of the channel at the pilot subcarriers using least square (LS) algorithm and interpolate the channel response at the data subcarriers. Then equalize the data to mitigate the multipath distortion.
5) Symbol detection and decoding: QPSK modulation and convolutional coding is used in the paper. After channel equalization, de-map and decode the data, obtain the recovered bits.

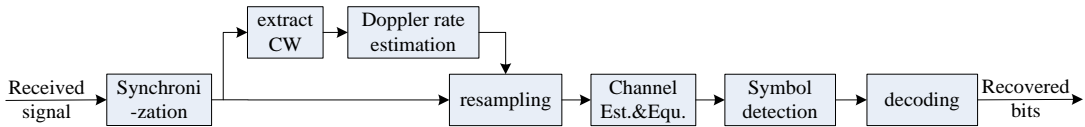

Fig. 2: Receiver diagram without time reversal processing.

### 3.2 Receiver with passive time reversal processing

Passive time reversal requires a probe signal sent prior to the data packet. PTR uses the received probes which are time-reversed and convolved with the data packet. In the procedure, an interference of the reversal probe is introduced. The output therefore needs to be convolved with the transmitted probe signal. Fig. 3 depicts the receiver diagram with passive time reversal processing. After resampling, the receiver extracts and time reverses the channel distorted probe signal; then convolves with the resampled packet and the transmitted probe signal. The procedure

can be expressed as

$$r'(t) = r(t) * p_r^*(-t) * p(t) \quad (12)$$

where * denotes convolution operation. *r(t)* is the received signal after resampling, *p(t)* and *p_r(t)* are the transmitted and received probe signals. *r(t)* and *p_r(t)* can also be expressed by the convolution

$$r(t) = s(t) * h(t) + n_1(t) \quad (13)$$

$$p_r(t) = p(t) * h(t) + n_2(t) \quad (14)$$

where $n_1(t)$ and $n_2(t)$ are the additive noise. s(t) is the transmitted data packet. Hence, the output after PTR processing becomes

$$\begin{aligned} r'(t) &= r(t) * p_r^*(-t) * p(t) \\ &= [s(t) * h(t) + n_1(t)] * [p^*(-t) * h^*(-t) + n_2^*(-t)] * p(t) \\ &= s(t) * h(t) * h^*(-t) * p^*(-t) * p(t) + n_3(t) \\ &= s(t) * h_{PTR}(t) + n_3(t) \end{aligned} \quad (15)$$

where

$$h_{PTR}(t) = h(t) * h^*(-t) * p^*(-t) * p(t) \quad (16)$$

$$n_3(t) = n_1(t) * p^*(-t) * h^*(-t) * p(t) + s(t) * h(t) * n_2^*(-t) * p(t) + n_1(t) * n_2^*(-t) * p(t) \quad (17)$$

Here, we define $h_{PTR}(t)$ as the passive time reversal channel which is treated as the final multipath channel the signal passes through after PTRP. $n_3(t)$ is the additive noise after PTRP. In the procedure, the received probe signal should contain all multipath information, so that the receiver could collect the energy of all multipath. But this does not mean that the longer is the better. Long received probes will introduce more noise.

From (16), we can observe that PRT channel is not an ideal $\delta$ function. Although the main lobe concentrates all multipath energy, the sidelobes still exist which introduce ISI. In a multiple receiver system, the interference caused by sidelobes could be reduced by using a dense array. For a single receiver, the interference cannot be avoided. Sidelobe levels depend on the multipath and the sidelobe levels of the probe signal's auto-correlation.

From (15)-(17) we can see that PRTP introduces mainly two types of interference: i) the self-generated interference caused by sidelobe of probe signal's auto-correlation (see (16)); ii) the cross-noise term introduced by the noise contained in the received probes (see(17)). In order to reduce the effect of the first type of interference, the probe signal should have a good auto-correlation. In our system, we use the frame synchronization signal (LFM) as the probe signal which has a good auto-correlation and there is no need to transmit an extra probe signal. After PTRP, the receiver needs to re-synchronize the frame and then detect the symbol.

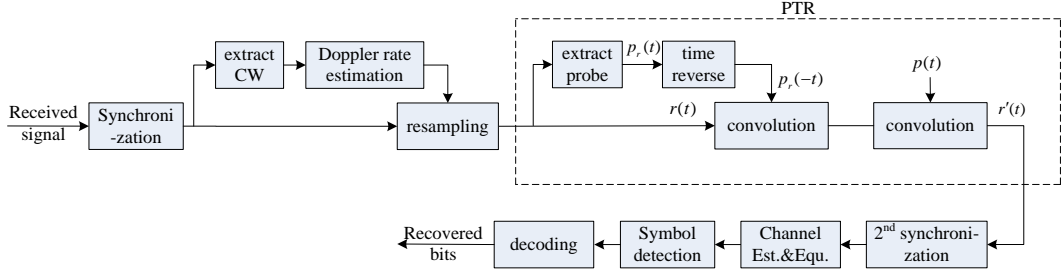

Fig. 3: Receiver diagram with passive time reversal processing.

### 3.3 Receiver with virtual time reversal processing

Since PTRP introduces interference caused by the probe signal, an alternative is to estimate the channel response. We then correlate the input signal with the reversed estimated channel response instead of the probe signal; this is the main idea behind virtual time reversal processing.

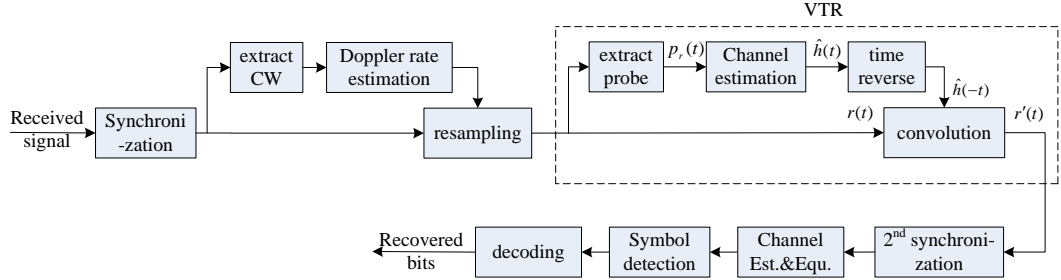

Fig. 4 shows the receiver diagram with virtual time reversal processing. After resampling, the receiver extracts the probe signal and estimates the CIR. Then it time reverses the estimated channel and convolves with the resampled received packet. Translating the procedure into equations, we have:

$$\begin{aligned} r'(t) &= r(t) * \hat{h}^*(-t) \\ &= [s(t) * h(t) + n_1(t)] * \hat{h}^*(-t) \\ &= s(t) * h(t) * \hat{h}^*(-t) + n_1(t) * \hat{h}^*(-t) \\ &= s(t) * h_{VTR}(t) + n_3(t) \end{aligned} \tag{18}$$

where

$$h_{VTR}(t) = h(t) * \hat{h}^*(-t) \tag{19}$$

$$n_3(t) = n_1(t) * \hat{h}^*(-t) \tag{20}$$

$\hat{h}(t)$ is the estimated CIR. Let us define $h_{VTR}(t)$ as the virtual time reversal channel and $n_3(t)$ as the additive noise after VTRP. Comparing (16), (17) with (19), (20), we can see that there is no interference of the probe signal for VTRP, and the performance depends on the estimated channel. If the channel estimation error is large, the performance of VTRP will greatly degrade due to the channel mismatch. Therefore, accurate channel estimation is needed for VTRP. In this paper, we compare two sparse channel estimation algorithms: matching pursuit (MP) and basis pursuit denoising (BPDN).

Another point that needs to be addressed is that after time reversal processing, the time reversal channel, (see (16) and (19)), is actually a non-minimum phase channel (path with the strongest power is not the first arrival path). Most of the time, we do not need to add a postfix if the first arrival path has the strongest power. For the non-minimum phase channel, a postfix is needed to mitigate the interference that is caused by the multipath before the path with strongest power. From (16) and (19), we can also find that, after TRP, the path with strongest power actually collects the energy of all multipaths which is the temporal compression, hence increasing the SNR. Therefore the TR system will gain more over rich multipath channels.

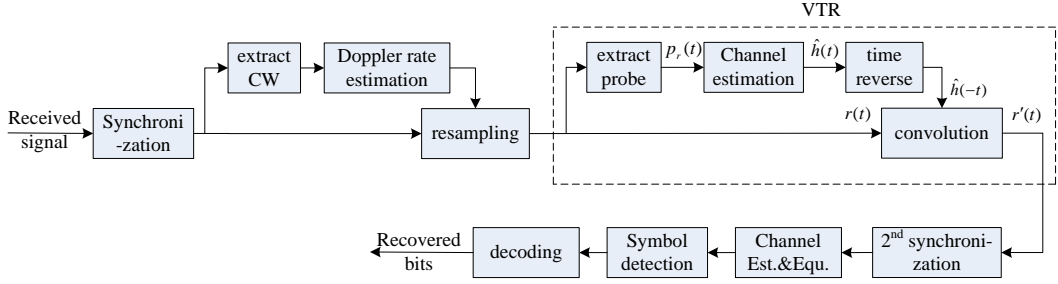

Fig. 4: Receiver diagram with virtual time reversal processing.

### 3.3.1 MP and BPDN Algorithms

VTRP requires knowledge of the CIR and the performance of VTRP mainly depends on the estimated CIR. In this paper, we compare two sparse channel estimation algorithms: matching pursuit (MP) and basis pursuit denoising (BPDN). We also propose an improved BPDN algorithm to make it more sophisticated in different noise backgrounds.

### 3.3.1.1 Matching Pursuit

Matching pursuit is a type of sparse approximation which finds sub-optimal solutions to the problem of an adaptive approximation of a signal in a redundant dictionary of functions. Consider the problem of solving for unknown signal **x** from the following,

$$\mathbf{y}=\mathbf{A}\mathbf{x} \tag{21}$$

where **A** is a $N \times M$ matrix and **y** and **x** are $N \times 1$ and $M \times 1$ vectors, respectively. Both **y** and **A** are assumed known. **A** is often called the dictionaries and the columns of **A** and $a_j$, for j=1,…,N, are called the atoms. A sparse solution $\hat{\mathbf{x}}$ may be viewed as the coefficient vector associated with the representation of **y** in terms of the $a_j$. The basic MP algorithm is an iterative procedure which can sequentially identify the dominant channel taps and estimate the associated tap coefficients. At each iteration, it selects one column of A that correlates best with the approximation residual from the previous iteration[23]. For details of MP algorithm the reader is referred to[23].

In our case the estimated CIR and the received probes can be viewed as **x** and **y**, respectively. A critical problem is how to build the dictionary. Some prior work has constructed the dictionary in frequency domain using pilots on different subcarriers[23][24]. In this paper, we use the preamble LFM as the probe to estimate the channel. The pulse-like autocorrelation of LFM probes are nearly orthogonal; we can therefore build the dictionary in time domain.

According to (8), after resampling, the received signal can be expressed as the convolution between the transmitted signal and channel impulse response. We rewrite (8) in the following format:

$$\underbrace{\begin{bmatrix} r(0) \\ r(1) \\ \vdots \\ r(N+L-1) \end{bmatrix}}_{\mathbf{r}} = \underbrace{\begin{bmatrix} s(0) & 0 & \cdots & 0 \\ s(1) & s(0) & \cdots & 0 \\ \vdots & \vdots & \ddots & \vdots \\ 0 & 0 & \cdots & s(N-1) \end{bmatrix}}_{\mathbf{s}} \underbrace{\begin{bmatrix} h(0) \\ h(1) \\ \vdots \\ h(L) \end{bmatrix}}_{\mathbf{h}} + \underbrace{\begin{bmatrix} v(0) \\ v(1) \\ \vdots \\ v(N+L-1) \end{bmatrix}}_{\mathbf{v}} \tag{22}$$

where N is the transmitted probe signal's length and $L$ is the maximum channel delay. $\mathbf{r}$ and $\mathbf{v}$ are $(N+L-1)\times 1$ vectors. $\mathbf{s}$ is a $(N+L-1)\times L$ matrix and $\mathbf{h}$ is a $L\times 1$ vector. Comparing (22) with (21), we can solve the channel estimation problem using the MP algorithm

$$\underbrace{\begin{bmatrix} r(0) \\ r(1) \\ \vdots \\ r(N+L-1) \end{bmatrix}}_{\mathbf{r}} = \underbrace{\begin{bmatrix} s(0) & 0 & \cdots & 0 \\ s(1) & s(0) & \cdots & 0 \\ \vdots & \vdots & \ddots & \vdots \\ 0 & 0 & \cdots & s(N-1) \end{bmatrix}}_{\mathbf{s=A}} \underbrace{\begin{bmatrix} \hat{h}(0) \\ \hat{h}(1) \\ \vdots \\ \hat{h}(L) \end{bmatrix}}_{\hat{\mathbf{h}}} \tag{23}$$

The matrix $\mathbf{s}$ can be viewed as the dictionary $\mathbf{A}$. $\mathbf{r}$ and $\hat{\mathbf{h}}$ are identical to $\mathbf{y}$ and $\mathbf{x}$ in (21). In the process of estimation, the received probes should be long enough to contain all multipath components.

### 3.3.1.2 Basis Pursuit Denoising

Basis pursuit denoising (BPDN) is an approach for solving the following convex optimization problem which is also referred to as $l_2$-$l_1$ problem

$$\min_{x \in \square^n} \frac{1}{2}\|\mathbf{y} - \mathbf{Ax}\|_2^2 + \tau \|\mathbf{x}\|_1 \tag{24}$$

where $\mathbf{y}$ is the observation vector, x is unknown vector, $\mathbf{A}$ is the dictionary matrix, and $\tau$ ($\tau \in \square^+$) is the regularization parameter. $\|\square\|_p$ stands for the $l_p$ norm (for $p \geq 1$), defined as $\|\mathbf{x}\|_p = \left(\sum_i |x_i|^p\right)^{1/p}$. Compared with the MP algorithm, BPDN adds a $l_1$ term to balance the sparsity ($l_1$ norm) and resilience to noise ($l_2$ norm). The regularization parameter controls the relative weight of the two terms. Sparse Reconstruction by Separable Approximation (SpaRSA) is a good solver for (24) proposed in [25]. SpaRSA repeatedly evaluates simple so-called *soft threshold* functions that transparently clip small entries in the real or complex coefficient vector to exactly zero[20]. In this algorithm, the parameter $\tau$ should be carefully chosen which could affect the convergence speed and the estimation accuracy. $\tau$ should be smaller than $\|\mathbf{A}^T\mathbf{y}\|_\infty$, otherwise, the unique solution to (24) is the zero vector[27]. In [25] the authors propose an adaptive continuation algorithm to choose the sequence of $\tau$ values but the initialized value is not specified. In the simulation, $\tau = 0.1\|\mathbf{A}^T\mathbf{y}\|_\infty$ and $\tau = 0.001\|\mathbf{A}^T\mathbf{y}\|_\infty$ were set to estimate the spike signal when the noise variance is $10^{-4}$ and zero in [22]. We propose a method on how to set the regularization parameter $\tau$ to balance the sparsity and the noise. $\tau$ is initialized by $\tau = 0.1*\sigma*\|\mathbf{A}^T\mathbf{y}\|_\infty$, where $\sigma$ is normalized noise variance (the signal power is 1). The details of the algorithm can be found in[29]. The proposed improved algorithm of SpaRSA could well balance the sparsity and the noise; therefore it improves channel estimation performance with different SNRs. The dictionary and observation vector are the same as that mentioned in Section 3.3.1.1.

## 4  Simulation results

In simulation, the signal is modulated by QPSK and encoded by 1/2 rate convolutional code. The bandwidth of the OFDM signal is 4kHz~12kHz and the sampling rate is 48kHz. 8192 points FFT is implemented for OFDM modulation which leads a 5.85Hz subcarrier spacing. A symbol duration lasts for 170.7ms. Pilot subcarriers are uniformly and even spaced distributed over the whole subcarriers. The cyclic prefix/postfix and the guard intervals between two signals last for 50ms which are longer than the simulation channel. LFM signal lasts for 40ms and the bandwidth is the also 4kHz~12kHz.

The multipath channel used in the simulation is generated by BELLHOP[28]. The measured sound speed profile of Songhua Lake, China in September is used in the simulation, shown in Fig. 5.The transmitter and receiver depth are 6m and 5.8m, respectively. The distance between two nodes is 2km. The simulated multipath channel is shown in .Fig. 6.

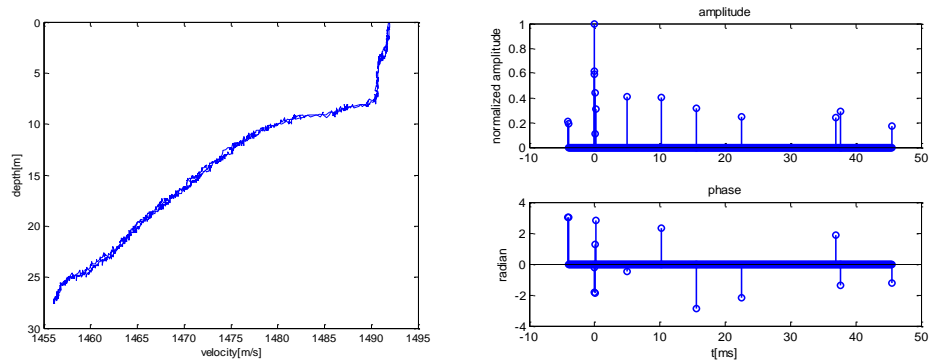

Fig. 5:Sound speed profile of Songhua lake.Fig. 6:Simulated multipath channel.

Figs.7 and 8 show the bit error rate (BER) for different pilot spacing.  A pilot spacing of 3 means that for every four subcarriers there is 1 pilot subcarrier and 3 data subcarriers. The figures compare the uncoded and coded BER of system without TRP, with TRP (PTRP, VRTP) and with fully known channel state information (CSI). For the known CSI simulation, there is no time reversal pre-process and zero forcing (ZF) equalizer is used. The processing procedure of the receiver is introduced in Section 3. In VRTP, we compare two channel estimation algorithms, MP and BPDN.

In general, the performance of the system with TRP (TR-OFDM)is much better than that without any TRP (No-TR-OFDM). TRP compresses the channel and collects the multipath energy which brings the processing gain. While, for the TR system, e.g. PTR, if no postfix is added, the uncoded BER may be even worse compared with the system without TRP. The coded BER of VTRP improves significantly compared to that of PTRP and is close to the BER when the receiver fully knows the channel. VTRP with BPDN channel estimation is a little better than using MP channel estimation. The simulation results reveal that the proposed VTR effectively reduces the interference introduced by the probe signal, and two channel estimation algorithms work well. The pool experiment in next section will further validate the proposed VTR performance.

There is still a gap between VTRP and known-CSI for uncoded BER even though the channel estimation error is small. This gap is generated by the interference caused by the sidelobes of the time reversal focus which is hardly reduced in single receiver system.

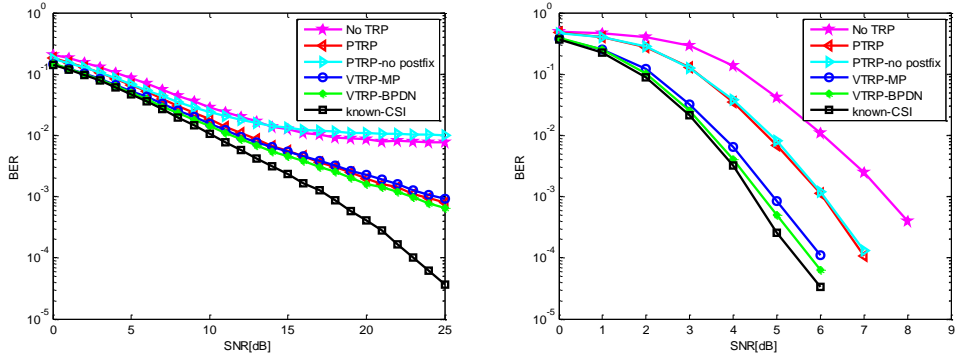

Fig. 7 (a) Uncoded BER, pilot spacing=3 Fig. 7(b) Coded BER, pilot spacing=3

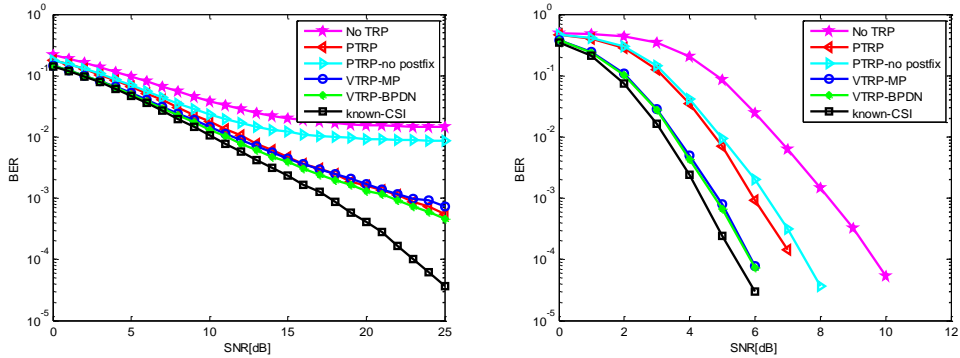

Fig. 8 (a): Uncoded BER, pilot spacing=5. Fig. 9 (b): Uncoded BER, pilot spacing=5.

## 5 Experimental results

The experiment was conducted in a pool with $45m \times 6m \times 5m$ size. The surrounding of the pool is wedge absorber with high absorption coefficient within the experimental bandwidth range and the bottom is sand. The cyclic prefix and postfix last for 20ms. The convolutional code is not used in the experiment. The other parameters are the same as that in the simulation. In the experiment, we use a laptop to transmit and receive the signal. The transmitted source is a picture with 80,000 bits. We change the transmit power to obtain different received SNRs. The pool channel impulse response estimated by BPDN is shown in Fig.9. Maximum channel delay spread is approximately 15ms.

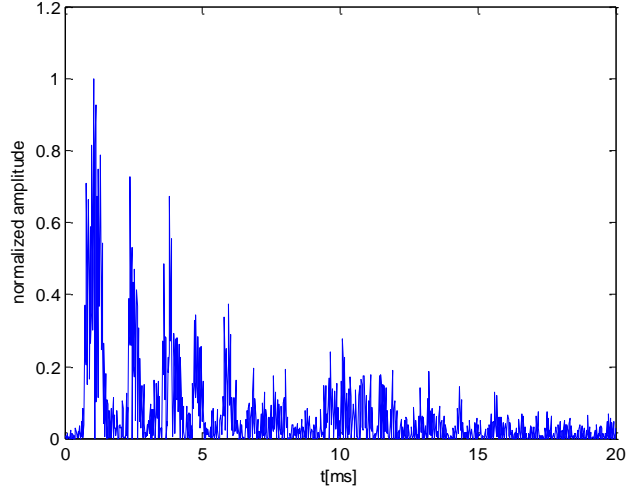

Fig. 9: Pool channel impulse response

The BER performances are shown in Fig.10(a) and Fig.11(a) for different pilot spacing. Input SNR (ISNR) is calculated in the time domain by comparing the received signal strength to the noise strength. The experimental results are consistent with the simulation results. TR-OFDM performs better than No-TR-OFDM and the gap is more obvious when the pilot spacing is large.

For better comparison, we use the effective signal-to-noise-ratio (ESNR) to evaluate the system performance which is used as a performance metric for mode switching in AMC-OFDM system in[21]. Different from the input SNR, ESNR accounts for the channel estimation error which can describe the processing gain better. ESNR is calculated as:

$$ESNR = \frac{E_{k \in S_D}\left[\left|\hat{H}[k]s[k]\right|^2\right]}{E_{k \in S_D}\left[\left|z[k] - \hat{H}[k]s[k]\right|^2\right]} \quad (21)$$

where $S_D$ is the ensemble of the data subcarriers, $\hat{H}[k]$ is the estimated channel response of the $k$-th subcarrier in frequency domain, $z[k]$ is the frequency observation at subcarrier $k$, and $s[k]$ is the transmitted symbols on subcarrier $k$.

Fig.10(b) and Fig.11(b) show the ESNRs with different processing. VTRP with MP channel estimation (VTRP-MP) improves little compared with PTRP when ISNR is small and gains about 0.5 dB with high ISNR. While VTRP with BPDN channel estimation (VTRP-BPDN) gains about 0.5-1.0dB compared with PTRP. This gain is coming from accurate channel estimation and reduction of interference of the probe signal's auto-correlation. VTRP which estimates the channel to reduce the interference introduced by probe signal gets better performance than PTRP. For two channel estimation algorithm, BPDN performs a little better than MP especially with low SNRs.

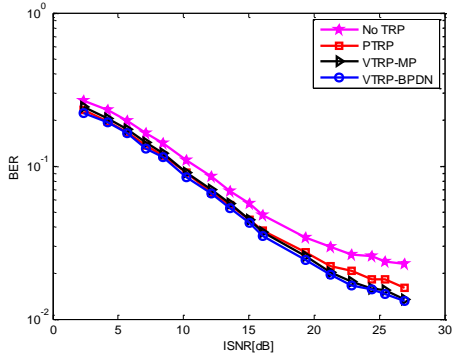 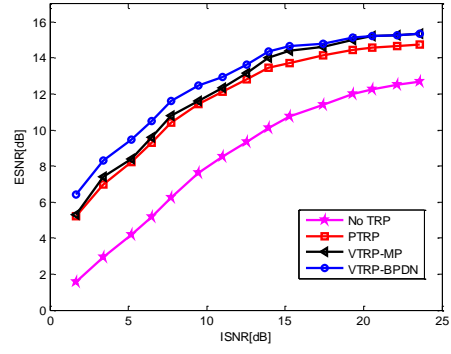

Fig. 10(a):BER performance, pilot spacing=3.    Fig. 10(b): ESNR, pilot spacing=3.

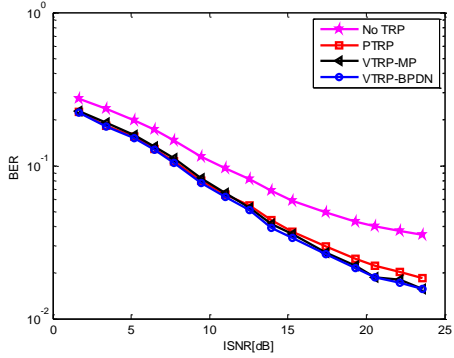 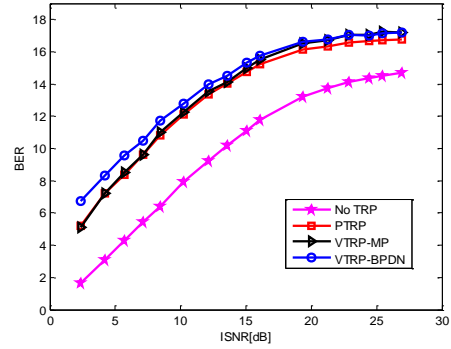

Fig. 11(a): BER performance, pilot spacing=5.    Fig. 11(b): ESNR, pilot spacing=5.

## 6  Conclusion

In this work, we examined the single sensor OFDM demodulation using virtual time reversal processing with two channel estimation algorithms in underwater communications. From the simulation and experimental results, we can discover that TRP-OFDM performs much better than No-TRP-OFDM over time-invariant channels. This is due to TRP compressing the received signal in time domain and collecting the entire multipath which enhances the received SNR. Our further results prove that the proposed VTRP performs better than PTRP and this gain is brought about by accurate channel estimation. VTRP removes the noise and the interference brought by the probe signal compared with PTRP. The performance of VTRP is mainly decided by the channel estimator. We compared two channel estimation algorithms: MP and BPDN. BPDN performs a little better than MP especially in low SNR.

**Acknowledgments**

We thank Allen Li for his careful edition to improve the readability of the manuscript.
This work was supported by the National Natural Science Foundation of China under Grant No.1127407, National High Technology Research and Development Program of China under Grant No. 2009AA093601-2 and the Underwater Acoustic Technology Laboratory Foundation of China under Grant No. 9140C200801110C2004.